# Negative optical spin torque wrench of a non-diffracting non-paraxial fractional Bessel vortex beam

F.G. MITRI*

*Corresponding author: F.G.Mitri@ieee.org

*Abstract* – **An absorptive Rayleigh dielectric sphere in a non-diffracting non-paraxial fractional Bessel vortex beam experiences a spin torque. The axial and transverse radiation spin torque components are evaluated in the dipole approximation using the radiative correction of the electric field. Particular emphasis is given on the polarization as well as changing the topological charge $\alpha$ and the beam's half-cone angle. When $\alpha$ is zero, the *axial* spin torque component vanishes. However, when $\alpha$ becomes a real positive number, the vortex beam induces *left-handed* (negative) axial spin torque as the sphere shifts off-axially from the center of the beam. The results show that a non-diffracting non-paraxial fractional Bessel vortex beam is capable to induce a spin reversal of an absorptive Rayleigh sphere placed arbitrarily in its path. Potential applications are yet to be explored in particle manipulation, rotation in optical tweezers, optical tractor beams, the design of optically-engineered metamaterials to name a few areas.**

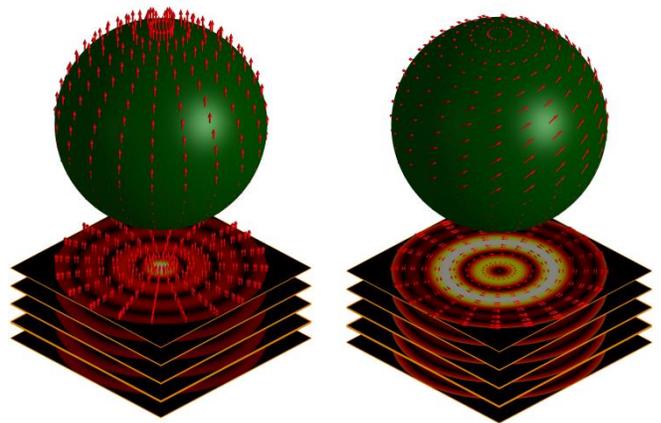

**Fig. 1.** The left panel shows the computational plot for the intensity vector field on a sphere with stereographic projection superimposed on the cross-sectional profile of a zeroth-order (non-vortex) Bessel beam. The right panel corresponds to the case of a first-order Bessel vortex beam.

The optical radiation torque [1] which arises from the transfer of angular momentum of light to a (lossy) dielectric sphere has received significant attention in optical levitation [2-4] and particle rotation applications in atomic physics [5, 6], cell biology [7] and other areas [8, 9]. In common optical laser beams, and under some conditions related to the particle absorptive properties, a *spin* torque causing the particle to rotate around its center of mass arises, in addition to an orbital component causing the particle to rotate around the beam's axis of wave propagation [1]. Note that the orbital torque component exerted on the sphere vanishes at the center of the beam. Moreover, the sphere should be absorptive in order to experience a *spin* torque [1]. Various standard types of laser beams have been suggested and investigated for this purpose [10, 11]. Examples of beams also include optical vortices of Gaussian-like [12, 13] and nondiffracting beams [14].

A particular kind of vortices with fractional topological charge (or order) has received increasing attention due to their special features and potential use in applications ranging from optical manipulation [15, 16], quantum entanglement [17], digital spiral imaging [18] among other topics. Typically, such types of vortices [15, 16] exhibit a diffractive slit opening while they propagate so the beam's cross-section becomes asymmetric.

In contrast, another class for factional vortex beams (which can be realized experimentally using blazed-phase hologram encoded in a programmable liquid crystal display illuminated with a He-Ne laser [19]) and displays *limited-diffracting* features during propagation exists [19, 20]. The physically-realizable apodized beam carrying finite energy preserves the nondiffracting propagation property. It is also known as a high-order Bessel vortex beam of fractional type $\alpha$ (HOBVB-F$\alpha$) [21-23]. A HOBVB-F$\alpha$ with fractional order $\alpha$ connects standard nondiffracting Bessel (vortex) beams (Fig. 1) of successive integer order in a smooth transition. It also generates individual vortices in the (lateral) plane perpendicular to the axis of wave propagation corresponding to a cross-section of the incident nondiffracting beam [19]. This specific feature provides the impetus to analyze the optical radiation *spin* torque exerted on a small dielectric absorptive sphere from the standpoint of the counter-intuitive "negative" optical radiation torque generation, meaning that the sphere would rotate around its center of mass in opposite handedness of the beam's angular momentum [24]. Note that this phenomenon has been originally observed from the standpoint of *acoustical* radiation spin torque theory using Bessel vortex beams [25]. The aim here is to suggest the optical HOBVB-F$\alpha$ as a potential candidate for the generation of a left-handed (negative) optical *spin* torque for applications yet to explore in particle manipulation and rotation using optical tweezers and tractor beams. Since the optical *orbital* torque component does not contribute to the rotation of the sphere around its center of mass, it will not be considered here. This topic, however, will be the subject of a forthcoming investigation.

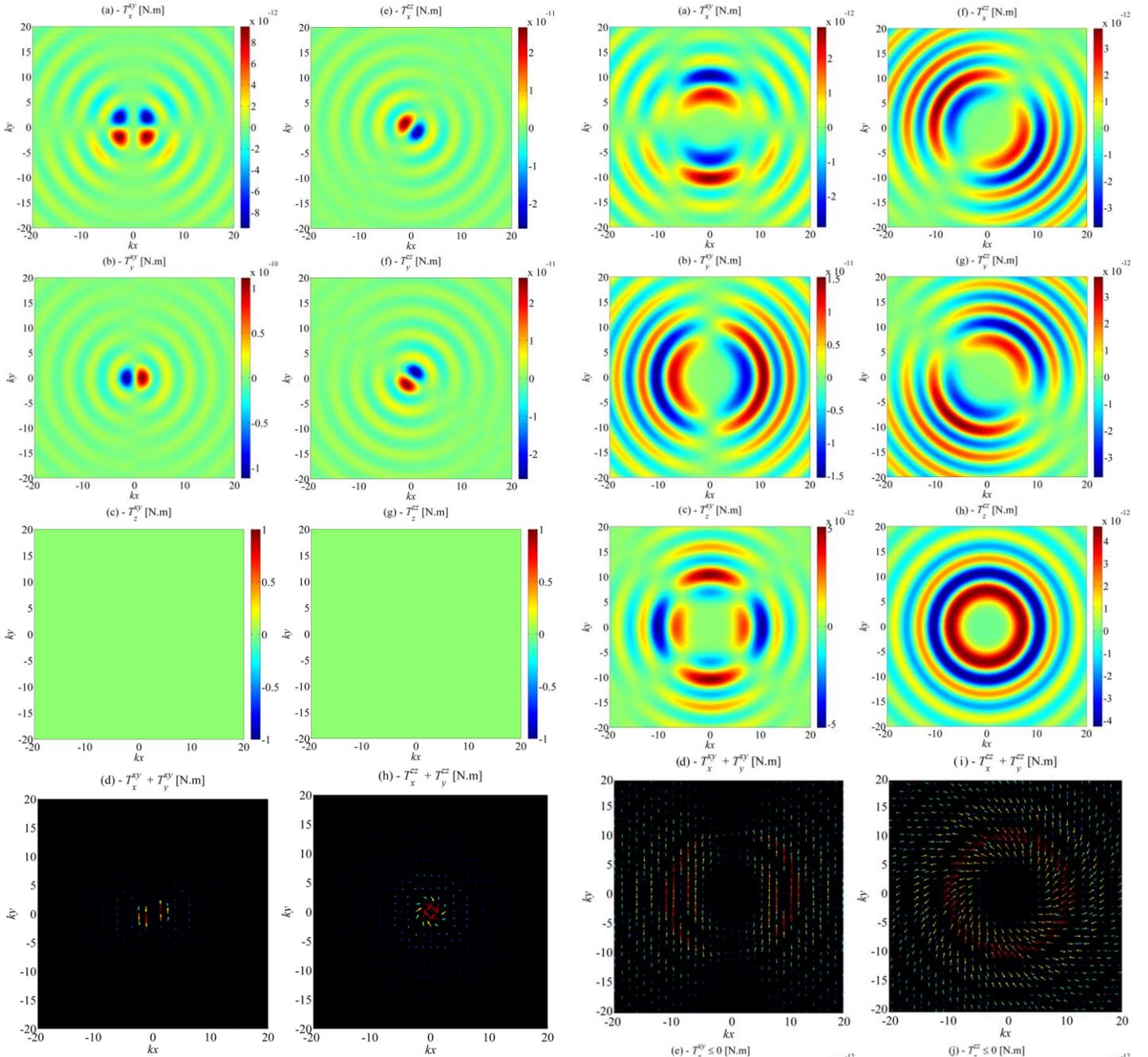

**Fig. 2.** The plots for the transverse radiation torque components for a zeroth-order Bessel beam where $\alpha = 0$, $\beta = 45°$, and $ka = 0.25$.

**Fig. 3.** The same as in Fig. 2 but $\alpha = 5$ and $\beta = 45°$. Note that the *axial* components $T_z^u$ of the torques are non-zero.

The analysis is started by considering the dipole approximation with the *modified* particle polarizability [26], which accounts for the radiative nature of the external field when it varies in time. The expression for the *spin* torque [27] is therefore given as,

$$\mathbf{T}^u\big|_{spin} = \frac{1}{2}|\alpha_e|^2 \Re\left\{\frac{1}{\alpha_e^{0*}} \mathbf{E}^u \times \mathbf{E}^{u*}\right\}, \quad (1)$$

where the modified electric polarizability is $\alpha_e = \alpha_e^0\left[1 - i\left(\frac{k^3 \alpha_e^0}{6\pi\varepsilon_0}\right)\right]$, $\alpha_e^0 = 4\pi\varepsilon_0 a^3(\varepsilon-1)/(\varepsilon+2)$, the superscript * denotes a conjugate of a complex number, $a$ is the radius of the sphere radius, $k$ is the wavenumber and $\varepsilon$ is the relative complex permittivity coefficient, corresponding to the ratio of the particle permittivity to that of the surrounding medium with permittivity $\varepsilon_0$. The spin torque can be evaluated directly from the expressions of the electric field components. In a Cartesian system of coordinates, where a *transverse* polarization scheme has been considered for the vector potential from which the electric (and magnetic) field can be derived, the electric field components are expressed as [21],

$$E_x^{xy} = \frac{1}{2}E_0 \sum_{m=-\infty}^{+\infty}\left\{\begin{array}{l} i^{(\alpha-m)}\text{sinc}(\alpha-m)\exp[i(k_z z + m\phi)] \\ \times\left[\left(1 + \frac{k_z}{k} - \frac{k_r^2 x^2}{k^2 R^2} + \frac{m(m-1)(x-iy)^2}{k^2 R^4}\right)J_m(k_r R)\right. \\ \left. - \frac{k_r(y^2 - x^2 - 2imxy)}{k^2 R^3}J_{m+1}(k_r R)\right] \end{array}\right\}, \quad (2)$$

$$E_y^{xy} = \frac{1}{2}E_0 xy \sum_{m=-\infty}^{+\infty}\left\{\begin{array}{l} i^{(\alpha-m)}\text{sinc}(\alpha-m)\exp[i(k_z z + m\phi)] \\ \times\left[\left(\frac{m(m-1)[2+i(x^2-y^2)/(xy)] - k_r^2 R^2}{k^2 R^4}\right)J_m(k_r R)\right. \\ \left. + \frac{k_r[2+im(y^2-x^2)/(xy)]}{k^2 R^3}J_{m+1}(k_r R)\right] \end{array}\right\}, \quad (3)$$

$$E_z^{xy} = \frac{1}{2}iE_0 \frac{x}{kR}\left(1 + \frac{k_z}{k}\right)\sum_{m=-\infty}^{+\infty}\left\{\begin{array}{l} i^{(\alpha-m)}\text{sinc}(\alpha-m)\exp[i(k_z z + m\phi)] \\ \times\left[\left(\frac{m(1-iy/x)}{R}\right)J_m(k_r R) - k_r J_{m+1}(k_r R)\right] \end{array}\right\}, \quad (4)$$

where $\varepsilon$ is the permittivity of the medium of wave propagation, $E_0 = ikA_0$, $A_0$ is the vector potential amplitude, $R = \sqrt{x^2 + y^2}$, $k_r = k\sin\beta$, $k_z = k\cos\beta$ and $\beta$ is the half-cone angle of the beam. Thus, for this configuration, the superscript $u$ in Eq.(1) would denote $xy$ indicating the state of polarization of the vector potential.

A different scheme is also chosen, in which an axial polarization scheme is considered for which $u$ corresponds to $zz$. The expressions for the electric in the axial polarization case are given as [22],

$$E_x^{zz} = \frac{A_0}{2k}\sum_{m=-\infty}^{+\infty}\left\{\begin{array}{l} i^{(\alpha-m)}\text{sinc}(\alpha-m)\exp[i(k_z z + m\phi)] \\ \left[m\frac{(k_z - ik)}{(x-iy)}J_m(k_r R)\right. \\ \left. - k_r \frac{(k_z x + ky)}{R}J_{m-1}(k_r R)\right] \end{array}\right\}, \quad (5)$$

$$E_y^{zz} = \frac{A_0}{2k}\sum_{m=-\infty}^{+\infty}\left\{\begin{array}{l} i^{(\alpha-m)}\text{sinc}(\alpha-m)\exp[i(k_z z + m\phi)] \\ \left[m\frac{(k_z - ik)}{(y+ix)}J_m(k_r R)\right. \\ \left. + k_r \frac{(k_z x + ky)}{R}J_{m-1}(k_r R)\right] \end{array}\right\}, \quad (6)$$

$$E_z^{zz} = \frac{A_0}{2k}\sum_{m=-\infty}^{+\infty}\left\{\begin{array}{l} i^{(\alpha-m+1)}\text{sinc}(\alpha-m)\exp[i(k_z z + m\phi)] \\ \times(k-k_z)(k+k_z)J_m(k_r R) \end{array}\right\}. \quad (7)$$

Notice that when $\alpha$ is an integer number, the plot for $|E_x^{zz}|$ is a rotated version of $|E_z^{xy}|$ [23].

The analysis is illustrated by considering numerical computations for the spin torque Cartesian components based on Eq.(1). In the computations, $ka = 0.25$, $\varepsilon = 2.528 + 0.0318i$, and the spin torque components have been calculated in the transverse plane within the ranges $-20 \leq (kx, ky) \leq 20$ for different values of the half-cone angle $\beta$ (denoted also in the caption of each figure).

Panels (a)-(d) of Fig. 2 show the plots for the spin radiation torque components of a zero-order Bessel non-vortex beam (i.e. $\alpha = m = 0$) having $\beta = 45°$ for the transverse polarization case, while panels (e)-(h) correspond to those obtained in the axial polarization configuration. Notice that the axial spin torque component $T_z^u$ vanishes in this case as the beam does not form a vortex along the $z$-direction as shown in panels (c) and (g). Notice that a transverse optical spin torque is exerted on the dielectric sphere as shown in panels (d) and (h) of Fig. 2.

The effect of increasing the order of the HOBVB-F$\alpha$ on the spin torque components to an integer $\alpha = m = 5$ is shown in panels (a)-(j) of Fig. 3 for $\beta = 45°$. In this case, the sphere experiences an axial (in addition to the transverse) spin torque for both polarization schemes ($xy$ and $zz$). In the $xy$ polarization case, the transverse torque component $T_y$ is larger [see panel (b)] and dominates in the transverse plane as shown in panel (d). As the sphere departs from the center of the vortex beam, it experiences positive and negative torques depending on its position in the transverse plane. Panels (e) and (j) show the negative values for which the *axial* torque is negative. As shown in panel (i), the arrows change their angular directions from the clockwise to the anticlockwise directions, such that the counter-clockwise sense correlates with the *negative* torque values of panel (j). Note also the positions in the lateral plane for which $T_z$ vanishes; i.e. it crosses zero from positive to negative values and vice versa. Furthermore, since the beam possesses a null in amplitude (or intensity) at its center so that no vortex can be formed along the direction of wave propagation $z$, all the components of the torque also vanish in the central region, as shown in all panels.

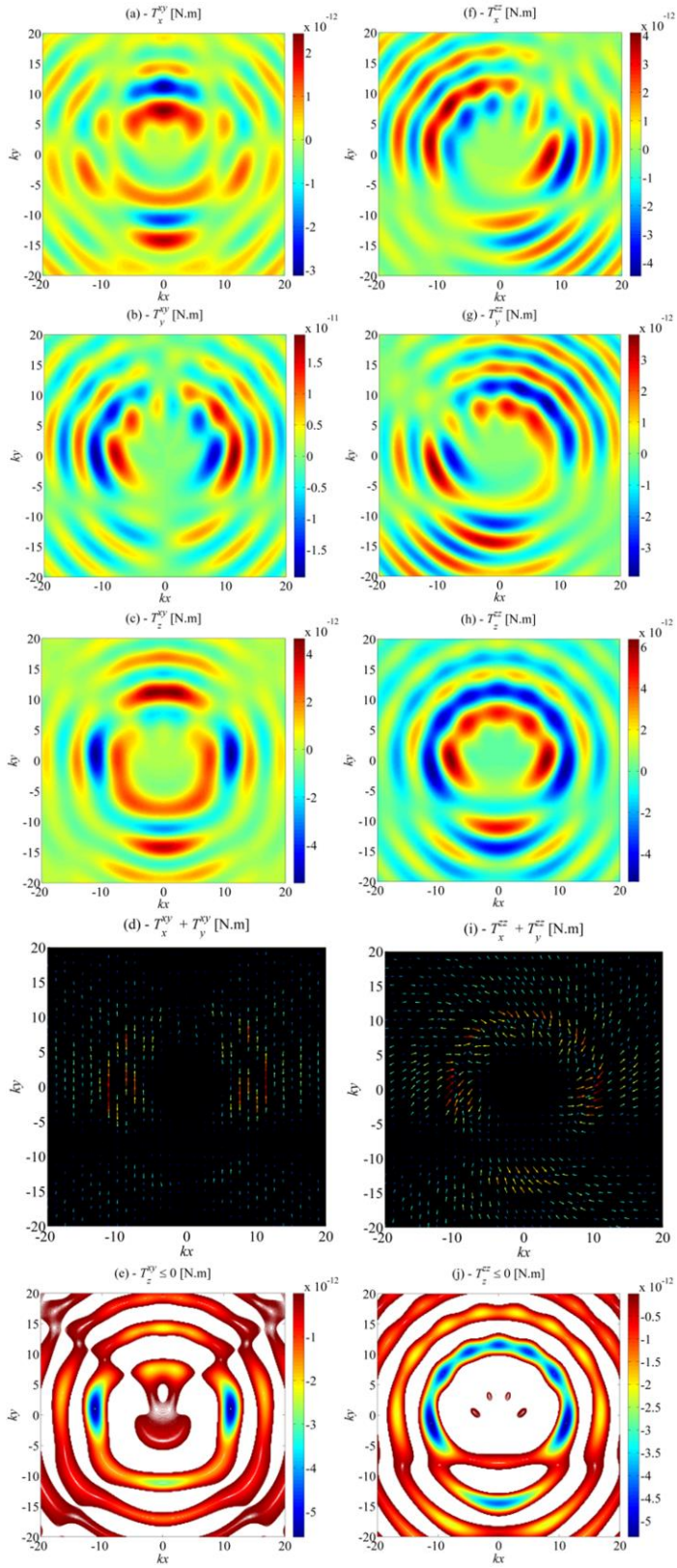

**Fig. 4.** The same as in Fig. 3 but $\alpha$ is fractional and equals 5.5, and $\beta = 45°$.

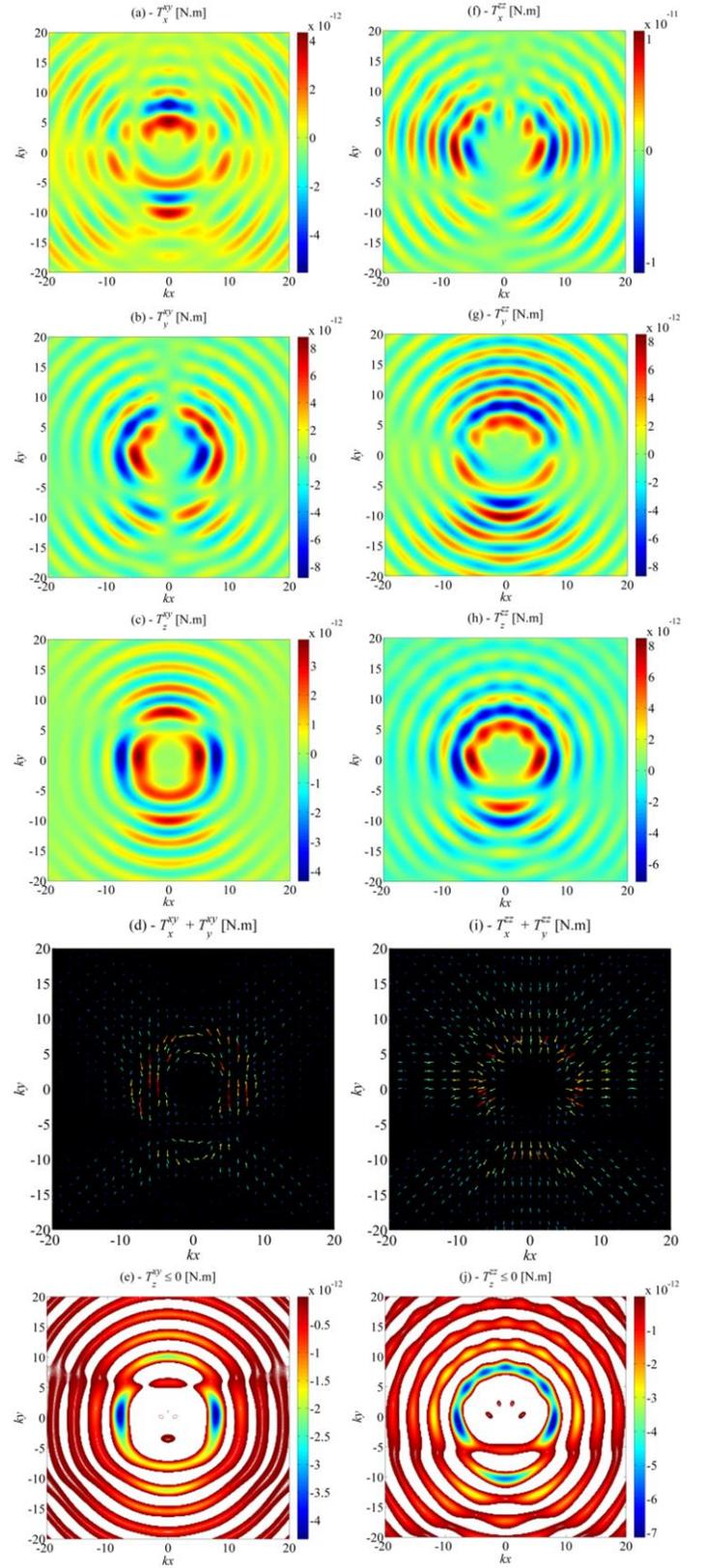

**Fig. 5.** The same as in Fig. 4 but $\alpha = 5.5$ and $\beta = 85°$.

When $\alpha$ becomes fractional (= 5.5), singularity features appear in the plots for the torque components, as shown in the panels of Fig. 4 for $\beta$ = 45°. Panel (e) shows that the axial torque becomes negative at the center of the beam in the $xy$-polarization state, while panel (j) indicates that the left-handed torque only occurs in the vicinity of the central region for the $zz$-polarization case. The circular symmetry observed previously in panel (i) of Fig. 3 is now broken in panel (i) of Fig. 4.

To further investigate the wide-angle effect of the beam, the half-cone angle $\beta$ is increased to 85°, which corresponds to a HOBVB-F$\alpha$ with a large conical-focusing half-angle. This has an influence on the left-handed axial spin torque components as shown in panels (e) and (j) of Fig. 5. Moreover, panel (d) shows an interesting feature of the transverse spin torque $T_x + T_y$ such that an anti-clockwise vortex-like feature is manifested below the central axis of the beam. One also notices that the transverse spin torque in panel (d) alternates from the clockwise to the anti-clockwise sense of rotation as the sphere departs from the beam's center. This panel shows that depending on its position in the transverse plane, the sphere can experience several individual lateral left-handed or right-handed spin torques within a single beam's cross-section when $\alpha$ is a fractional number.

In the present investigation, a single non-paraxial HOBVB-F$\alpha$ has been considered with emphasis on the generation of left-handed axial spin torques. Nonetheless, in applications involving optical tweezers and particle levitation and rotation, two (or multiple) counter-propagating HOBVBs-F$\alpha$ [23] could be used, which is expected to enhance the trapping capabilities. The analysis can be directly extended to investigate the generation of negative spin torques using stationary (i.e. counter-propagating) optical beams.

Another topic of interest concerns the extension to the case of a magneto-dielectric (semi-conducting) sphere, which presents unique scattering properties such that the forward scattering or the backscattering can be enhanced or reduced depending on its physical properties [28]. In this case, the magnetic field would contribute to the generation of a magnetic spin torque [29, 30] so that the total (i.e. electric + magnetic) left-handed (negative) spin torque rotating the magneto-dielectric sphere around its center of mass may be enhanced or reduced. This topic will be the subject of a forthcoming research.